\documentclass[epj]{svjour}
\usepackage{graphics}
\begin{document}
\title{ Short-range NN-properties in the processes $pd\to dp$
 and $pd\to (pp)n$}
\author{Yu.N. Uzikov\inst{1,2}, {V.I. Komarov}\inst{2}, F.Rathmann\inst{3}, 
 \and H. Seyfarth\inst{3} 
}                     
\offprints{Yu.N. Uzikov, Laboratory of Nuclear Problems, JINR,
 Dubna, Moscow region, Russia  141980.}          
\institute{Kazakh National University, 480078 Almaty, Kazakhstan
\and Laboratory of Nuclear Problems, Joint Institute for Nuclear Research,
 Dubna, Russia 141980
\and Institut f\"ur Kernphysik, Forschungszentrum J\"ulich,
52425 J\"ulich, Germany 
}
\date{Received: date / Revised version: date}
%

\abstract
{The data from the first measurement, performed at ANKE/COSY, of the
unpolarized cross section of the reaction $pd\to (pp)n$ in the
kinematics of backward elastic $pd\to dp$ scattering at 
proton-beam energies between 0.6 and 1.9 GeV are analyzed in a 
phenomenological approach. The $pd\to (pp)n$ data
 and the triplet cross section of the reaction $pd\to (pn)p$,
 calculated from the $pd\to dp$ data on the basis of
 the F\"aldt-Wilkin extrapolation,  are used here
to derive the ratio $\zeta$ of the singlet production matrix element
 squared to the triplet one. 
This ratio, defined in our earlier analysis of
$pd\to (pn)p$ data
 in a largely model-independent way,
 depends on   
  the dynamics  of the $pd$ interaction. We find 
 here  $\zeta\approx 0.02$ and show that the smallness of this value
may point toward softness of the
 deuteron at short NN distances.
} 
\authorrunning{Yu.N. Uzikov et al.} 
 \titlerunning{Short-range NN-properties}
\maketitle
\section{Introduction}
\label{intro}

 The  structure of the lightest nuclei at short distances in the nucleon
overlap region $r_{NN}< 0.5 $ fm, {\it i. e.} at high relative momenta
 $q_{NN}\sim 1/r_{NN}> 0.4 $ GeV/c between the nucleons, is 
a fundamental problem of nuclear  physics. The structure can be  
tested by electromagnetic probes at high transferred momenta.
 However, a self-consistent picture 
of electro- and photo-nuclear processes is not yet developed
 due to the unknown strength of the meson-exchange currents. 
Hadron-nucleus collisions can give important independent informations.
 The theoretical analysis of hadron processes is complicated
 by initial and final state interactions and the excitation/de-excitation
 of  nucleons in intermediate states. For instance, a large contribution
 of the double $pN$ scattering with excitation of the $\Delta(1232)$-
 resonance was found in proton-deuteron backward elastic scattering
 $pd\to dp$ at $\sim 0.5$ GeV \cite{cwilkin69,kls,bdillig,imuz88,uz98}.
 At higher beam energies the role of heavier baryon resonances is expected
 to increase. Unfortunately, these contributions are not well controlled in theory
 due to the rather poor information about the $pN\rightleftharpoons NN^*$and 
 $pN\rightleftharpoons N\Delta$ amplitudes. To some extent these effects
 are taken into account in the one-pion exchange (OPE) model \cite{cwilkin69}
 with a virtual subprocess $pp\to d\pi^+$, but another important contribution,
 i.e. the one-nucleon-exchange (ONE) amplitude, cannot be included
 in this model.

 To minimize these complicating effects, it was proposed \cite{imuz90}
 to study the deuteron breakup reaction $pd\to (pp)n$ in the kinematics
 of backward  elastic $pd$ scattering. For low excitation energies
 $E_{pp} \leq 3$ MeV, the final $pp$ pair can be assumed to be in the
 $^1S_0$ spin singlet (isotriplet) state. This feature, in contrast to
 the $pd\to dp$ process, results in a considerable suppression of the
 $\Delta-$ (and $N^*)-$ excitation amplitudes by the isospin factor 1/3
 in comparison with the one-nucleon exchange (ONE). Recently it was shown
 \cite{uzzhetf} that the same suppression factor acts for a broad class
 of diagrams with isovector meson--nucleon rescattering in the intermediate
 state including the excitation of any baryon resonance. Furthermore,
 the node in the  half-off-shell $pp(^1S_0)$ scattering amplitude
  at the off-shell momentum $q\sim 0.4$ GeV/c results in remarkable
 irregularities in the spin observables and leads to a dip in the unpolarized
 cross section for the ONE mechanism \cite{imuz90,uz2002}. In the $pd\to dp$
 and $pd\to pX$ processes, the node in the deuteron S wave is hidden 
 by the large  contribution of the D wave. The irregularities of the
  observables allow new studies of (i) the commonly used potentials of 
 the NN interaction at short distances and (ii) possible contributions from
 $N^*$- exchanges \cite{kk} and exotic three-baryon states \cite{kls}.

 The first  data on the reaction $pd\to (pp)n$ at high beam energies
 $T_p=0.6 - 1.9$ GeV
 with forward emission
 of a fast proton pair of low excitation energy $E_{pp}$ were obtained 
at ANKE-COSY
\cite{vikomarov}. Using a
 largely model-independent approach based on the Migdal-Watson and the
 F\"aldt-Wilkin  FSI theory,  we discuss the relative strength of the measured
 singlet in comparison to the triplet channel. We present also the  results of
 a calculation of the deuteron  breakup cross section performed  within the
 main mechanisms of the $pd\to dp$ process \cite{uz98}.

\section{The FSI theory }
\label{sec:1}

At excitation energies around 1~MeV,  the $pd\to(pn)p$
cross section is strongly influenced by the $np$ FSI. The resulting
peak is well described by the Migdal-Watson formulae~\cite{Watson,GW}
 which take into account the nearby poles in the FSI triplet ({\it t})
 and singlet ({\it s}) $pn-$scattering amplitudes
\begin{equation}
\label{mwatson}
 d\sigma_{s,t} =FSI_{s,t}(k^2)\, K\, |A_{s,t}|^2.
\end{equation}
 Here $A_{s,t}$ is the production matrix element for the
 singlet and triplet state, $K$ is the kinematical factor,
 and $FSI_{s,t}$ is the  Goldberger-Watson factor \cite{GW}
\begin{equation}
\label{gwfsi}
 FSI_{s,t}(k^2)=\frac{k^2+\beta_{s,t}^2}{k^2+\alpha_{s,t}^2}.
\end{equation}
 Here k is the relative momentum in the $pn$ system at
 the excitation energy $E_{np}=k^2/m$, where $m$ is the nucleon mass.
The parameters $\alpha$ and $\beta$  are determined by the known 
 properties of the on-shell $NN$-scattering  amplitudes at low energies: 
  $\alpha_t=0.232$ fm$^{-1}$, $\alpha_s=-0.04$ fm$^{-1}$,
 $\beta_t=0.91$ fm$^{-1}$, $\beta_s=0.79$ fm$^{-1}$ \cite{machl}.
 Information on the $pd\to pnp$ mechanism and the off-shell
 properties of the {\it NN} system is contained in the  matrix elements
 $A_{s,t}$ and their ratio \cite{ukrs}
\begin{equation}
\label{zeta}
\zeta=\frac{|A_{s}|^2 }{|A_{t}|^2 }.
\end{equation}
 The $pd\to (pp)n$ data, obtained at ANKE/COSY \cite{vikomarov}, are presented
 as  cms cross sections
\begin{equation}
\label{2fold}
{\overline {\frac{d\sigma}{ d\Omega_n}}}=
 \frac{1}{\Delta \Omega_n}\int_0^{E_{max}}dE_{pp}\int\int
 m\frac{d^3\sigma}{dk^2 d\Omega_n} d\Omega_n,
\end{equation}
 integrated over $E_{pp}$ from 0 to 3 MeV and averaged over
 the neutron cms angle $\theta_n^*=172^\circ-180^\circ$, where
\begin{equation}
\label{3fold}
\frac{d^3\sigma}{dk^2 d\Omega_n}=\frac{1}{(4\pi)^5} \frac{p_n}{p_i}
\frac{k}{s\, \sqrt{m^2+k^2}}\frac{1}{2}\int\int d\Omega_{\bf k}
{\overline {|M_{fi}|^2}}
\end{equation}
 and $\Delta\Omega_n$ is the neutron solid angle. In Eq. (\ref{3fold})
 $p_i$ and $p_n$  are the cms momenta of the incident proton and the
 final neutron, respectively; $M_{fi}$ is the full matrix element of
 the reaction.
 Due to the isospin invariance the following relation holds
 in the singlet cannel
\begin{equation}
 \frac{d\sigma}{d\Omega^*}(pd\to {(np)}_sp)=\frac{1}{2}
\frac{d\sigma}{d\Omega^*} (pd\to {(pp)}_sn).
\label{isospin}
  \end{equation}  
 Using the F\"aldt-Wilkin extrapolation \cite{FW1} for the bound
 and the scattering S wave functions in the triplet state
 at short {\it pn} distances $r<1$~{\rm fm}, and taking into account
 the short-range character of the interaction mechanism, 
 the triplet cross section $pd\to (pn)_tn$ is obtained as
\begin{equation}
\label{triplet}
\frac{d\sigma_t}{d\Omega^*} =\frac{p_f}{p_i} \, f^2(k^2)
 \frac{d\sigma}{d\Omega^*}(pd\to dp),
\end{equation} 
 where 
\begin{equation}
\label{fsi}
f^2(k^2) = \frac{2\pi\,m}{\alpha_t(k^2+\alpha_t^2)}
\end{equation}
and $d\sigma/ d\Omega^*$ is the $pd\to pd$ cms cross section. After
 integration over $E_{pp}$ the triplet cross section (\ref{triplet})
 takes the form
\begin{equation}
\label{atriplet}
{\overline {\frac{d\sigma_t}{d\Omega^*}}}=
\frac{p_f}{p_i} Z \frac{ d\sigma}{d\Omega^*}(pd\to dp),
\end{equation} 
where
\begin{equation}
\label{z}
 Z=\frac{1}{2\pi}\frac{2}{\alpha_t}\left \{ k_{max}-\alpha_t 
 \arctan{\left(\frac{k_{max}}{\alpha_t}\right )} \right \}.
\end{equation} 
On the other side, the triplet (t) and singlet (s) cross sections
can be obtained by integration of Eq. (\ref{mwatson})  over $E_{pp}$.
$K$, $|A_s|^2$, and $|A_t|^2$, being very smooth functions of $E_{pp}$,
 can be assumed as constant.
 A ratio of the integrals $R=y_s/y_t$  can
be defined, where
\begin{equation}
\label{yst}
y_{s,t}=\int_0^{k^2_{max}}\,FSI_{s,t}(k)\,k\, dk^2.
\end{equation}
With this ratio one finally gets the singlet-to-triplet ratio $\zeta$ defined by
Eq. (\ref{zeta}) as
\begin{equation}
\label{zetacosy}
\zeta= \frac{1}{2\,R\, Z}
\frac{\frac{d\sigma}{d\Omega^*}(pd\to(pp)_sn)}
{\frac{ d\sigma}{d\Omega^*}(pd\to dp)}.
\end{equation}
It is obvious that $\zeta$ is not a direct ratio of the
$pd\to (pp)n $ and $pd\to dp$  cross sections, but contains also
   the additional  factor $1/({2\,R\,Z})$.

\section{Results and discussion}
For the numerical calculations of $\zeta$, defined by Eq. (\ref{zetacosy}),
we use the experimental data on the $pd\to dp$ cross section
\cite{berth} and the new ANKE/COSY data for the $pd\to (pp)n$ reaction 
\cite{vikomarov}. From the Eqs.(\ref{gwfsi}) and (\ref{yst}) and with the
 values for $\alpha_i$ and $\beta_i$, given above, we get $Z=0.101$ and
 $R=2.29$ for $E_{pp}^{max}=3$ MeV. With these numbers and assuming systematic
 uncertainties of 10\% for both the $pd\to dp$ cross section and the
 F\"aldt-Wilkin ratio (\ref{triplet}), we obtain $\zeta=(2.3\pm 0.5)$\%
 for 0.6, $\zeta=(1.6\pm 0.3)$\% for 0.7, and $\zeta=(2.1\pm 1.2)$\% for
 1.9 GeV beam energy. We find the surprising fact that $\zeta$ is constant
within the errors over the whole investigated beam-energy range from 0.6 to 1.9 GeV.

 The present, model-independent small result for $\zeta$ can be compared
 with $\zeta <5$\%, obtained recently \cite{ukrs} for the
 $pd\to (pn)p$ reaction data, measured  exclusively
 at 585 MeV and cms angle $\theta^*=92^\circ$ \cite{witten}. 
 A similar, small value for the singlet admixture of a few percent
  was found for the reaction $pp\to pn\pi^+$ in the FSI region at beam
 energies of 492 MeV \cite{abaev} and 800 MeV \cite{uzcw}, too.  The smallness
 of the singlet contribution in the $\Delta$-region of the reaction
 $pd\to (pn)p$ can be explained by dominance of the OPE mechanism 
 with the subprocess $pp\to (pn)_s\pi^+$ \cite{ukrs}.

 The singlet-to-triplet ratio can be estimated as
 $\zeta_{th}=R_S\times R_I\times R_X$.
 Here $R_S=1/3$ is the spin statistical factor. The isospin ratio $R_I$
 is 1 for the ONE and 1/9 for both the $\Delta$-mechanism and the
 vector meson--nucleon exchanges \cite{uzzhetf}. $R_X$ is the ratio of
 the spatial singlet and triplet amplitudes. For the  $\Delta$-mechanism,
 it reflects the difference of the $^1S_0$ and $^3S_1$ wave functions
 at $r < 1 $ fm. Since $\zeta$ according to Eqs. (\ref{mwatson}) and 
(\ref{gwfsi}) does not contain the FSI factor, $R_X$ is $\approx 1$ for
 the $\Delta$ amplitude. For the ONE it is $\approx 0.5$ due to the 
 contribution of the D wave in the triplet and its absence in  the singlet
 state.
 The calculation using Eqs.(\ref{3fold}) and (\ref{2fold}) with
 the Reid-Soft-Core (RSC) NN-potential \cite{rsc} gives
 $\zeta_{ONE}=5-8$\% for $T_p=1.4-1.9$ GeV  for the ONE which is 3 to 4 times
 the experimental value. 
 For the $\Delta$ mechanism this ratio
 $\zeta_{\Delta}=R_S\times R_I=1/27$ is
in better agreement with the experimental value.

 Calculations \cite{imuz88,uz2002} with use of the RSC NN potential and the 
 ONE+SS+$\Delta$ model \cite{kls,uz98}, including single pN scattering (SS), 
 describe the $pd\to (pp)n$ cross section data for the beam energies 0.6 and
 0.7 GeV as is seen in Fig. 1. In this region the $\Delta$ mechanism
 dominates due to the nearby minimum of the ONE cross section. At higher
 energies  the strong disagreement with the data is obvious. The data show
 no indication of the dip around 0.8 GeV, and for $T_p>1.3 $ GeV they are
 by a factor 2 to 4 below the prediction. A very similar discrepancy is observed for
 $pd\to dp$
 backward elastic scattering, as can be seen from the upper part of Fig. 1.
 Both the earlier $pd\to dp$ and the recent $pd\to (pp)n$ data show that in
 the until now used model the contribution by the ONE, dominating for
 $T_p>1.3$ GeV, is too strong. On the other side,
 a calculation, using only the $\Delta$
 mechanism with a cutoff momentum $\Lambda_\pi=0.7$ GeV/c allows one 
 to describe the $pd\to (pp)n$ data for $T_p>0.7$ GeV. This mechanism
does not involve such high-momentum components of the NN wave function
 as the ONE.
 A possible conclusion would be
 that the deuteron
 and the  $pp(^1S_0)$ system at short NN distances are softer than modeled by
 the RSC NN potential. 
As one should note, the assumption about softness of
 the deuteron at short NN-distances is supported 
 by (i) rather
 successful
 description of the $pd\to dp$ cross section within the OPE-model only
 \cite{kolybas,uz98}, and (ii) the strong disagreement between the ONE calculations
 and the $T_{20}$ data \cite{t20}. Detailed analyses of this conjecture are
 in progress.
   
\begin{figure}
\resizebox{0.5\textwidth}{!}{%
  \includegraphics{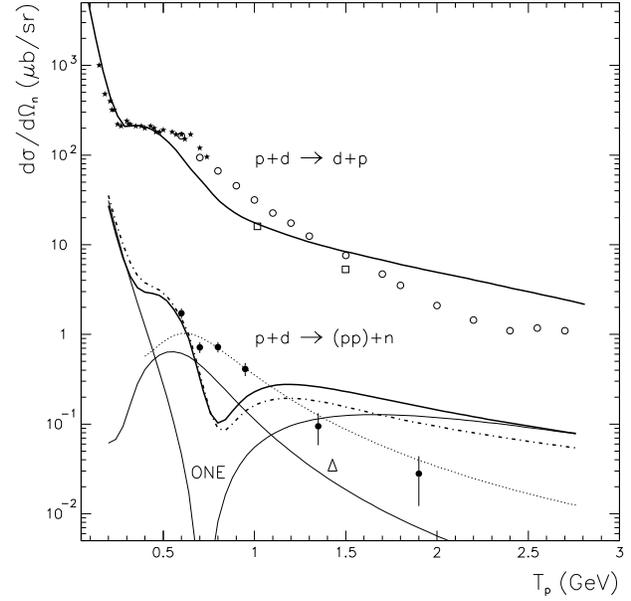}
 }
\caption{Experimental cross section of the reactions $pd\to dp$ \cite{berth}
  and $pd\to (pp)n$ \cite{vikomarov} as function of the beam energy in
 comparison with ONE+SS+$\Delta$ model calculations (thick full lines) 
 using the RSC potential  with cutoff momentum $\Lambda_\pi=0.53$ GeV/c
 in the $\pi NN$ and $\pi N\Delta$-vertices \cite{imuz88}. Inclusion of 
rescattering in the initial and final states, taken into account
 within the ONE(DWBA)+SS+$\Delta$ model \cite{uz2002}, yields the dashed-
dotted line. The dotted line results from a calculation using only the 
$\Delta$ mechanism with cutoff momentum $\Lambda_\pi=0.7$ GeV/c.}
\label{fig:1}       
\end{figure}
%
%
%
%
{\bf Acknowledgments.} This work was supported by the BMBF project
grant KAZ 99/001 and the Heisenberg-Landau program.

\end{document}